%% file: main.tex
\documentclass[runningheads]{llncs}

\usepackage[T1]{fontenc}
\usepackage{cite}
\usepackage{amsmath,amssymb,amsfonts}
\usepackage{algorithmic}
\usepackage{graphicx}
\usepackage{xcolor}
\usepackage{paralist}
\usepackage{subcaption}
\usepackage{url}
\usepackage{xspace}
\usepackage{enumitem}
\usepackage{booktabs}
\usepackage{multirow}
\usepackage[table]{xcolor}
\usepackage[subtle]{savetrees}

\def\BibTeX{{\rm B\kern-.05em{\sc i\kern-.025em b}\kern-.08em
    T\kern-.1667em\lower.7ex\hbox{E}\kern-.125emX}}

\newcommand{\fig}[1]{Fig.~\ref{fig:#1}}
\newcommand{\tab}[1]{Table~\ref{tab:#1}}
\newcommand{\s}[1]{Sec.~\ref{sec:#1}}

\newcommand{\fakeparagraph}[1]{\smallskip\noindent\emph{#1.}}
\newcommand{\sysname}{WASP\xspace}
\newcommand{\modeAot}{AOT\xspace}
\newcommand{\modeInt}{Default\xspace}

\newcommand{\noteam}[1]{\footnote{\textcolor{red}{Ale: #1}}}
\newcommand{\notedd}[1]{\footnote{\textcolor{blue}{Dario: #1}}}
\newcommand{\notemc}[1]{\footnote{\textcolor{teal}{Matteo: #1}}}

\renewcommand{\noteam}[1]{}
\renewcommand{\notedd}[1]{}
\renewcommand{\notemc}[1]{}

\begin{document}

\title{\sysname: A Configurable Framework for Portable Stateful Serverless
  Applications in the Edge-Cloud Continuum}

\titlerunning{\sysname: A Configurable Framework for Portable Stateful Serverless
  Applications}

\author{Matteo Cenzato\inst{1} \and
  Dario d'Abate\inst{1} \and
  Arianna Dragoni\inst{1} \and
  Giacomo Orsenigo\inst{1} \and
  Luca Tosetti\inst{1} \and
  Alessandro Margara\inst{1}}

\authorrunning{M. Cenzato et al.}

\institute{Politecnico di Milano, Milan, Italy \\
  \email{\{dario.dabate, arianna.dragoni, alessandro.margara\}@polimi.it}\\
  \email{\{matteo.cenzato, giacomo.orsenigo, luca1.tosetti\}@mail.polimi.it}}

\maketitle

\sloppy

\input{abstract}

\input{intro}
\input{background}
\input{design}

\input{eval}

\input{conclusions}

\bibliographystyle{splncs04}
\bibliography{biblio.bib}

\end{document}

%% file: abstract.tex
\begin{abstract}
  WebAssembly (WASM) is emerging as a lightweight alternative to containers
  for Function-as-a-Service (FaaS) across the edge-cloud continuum.
  However, existing WASM-based serverless platforms are tightly coupled to
  specific execution engines and predominantly designed for stateless
  workloads. This clashes with the heterogeneity of edge deployments, which
  demand support for stateful applications under diverse hardware and workload
  constraints.
  We introduce \sysname, a configurable framework that brings stateful serverless
  execution to the edge-cloud continuum. By abandoning monolithic
  architectures in favor of strictly decoupled, pluggable components, \sysname
  lets system administrators swap the WASM runtime and the datastore to fit
  available resources and application requirements, without altering
  application code. Configurable lifecycle and caching policies further enable
  fine-tuning for diverse non-functional requirements.
  Our experimental evaluation demonstrates that \sysname introduces negligible
  runtime overhead and, by swapping runtimes, datastores, and policies,
  exposes radically different memory and latency profiles, confirming its
  adaptability to the heterogeneous constraints of the edge-cloud
  continuum.
\end{abstract}

%% file: intro.tex
\section{Introduction}
\label{sec:intro}

Driven by the proliferation of data-intensive applications and IoT devices,
the edge-cloud continuum is rapidly emerging as a highly promising deployment
paradigm for modern distributed
systems~\cite{shi:edge-comp:iot-journal:2016,bittencourt:continuum:2025:compsci-review}.
In this context, the Function-as-a-Service (FaaS) programming model is gaining
significant traction~\cite{nastic:serverless-edgetocloud:internet-comp:2017}.
By allowing developers to structure applications as independent, event-driven
functions, FaaS entirely abstracts the underlying infrastructure management,
significantly simplifying the development lifecycle and enabling dynamic
workload placement across distributed nodes.
The traditional FaaS model is inherently stateless: function instances are
ephemeral entities created to handle specific requests and destroyed shortly
after, retaining no local context.
As a result, the standard architectural pattern for building complex
serverless applications involves leveraging external datastores, typically
offered as managed cloud services, to persist state across multiple function
invocations.

To support these applications directly at the edge while reducing execution
costs, WebAssembly (WASM) is positioning itself as an efficient alternative to
traditional container-based FaaS runtimes~\cite{hoque:Magazine:2022:WASMEdge}.
WASM provides a lightweight, memory-safe, and language-agnostic execution
sandbox, supported by a vast ecosystem of runtimes, each offering different
execution models (e.g., Ahead-of-Time compiled, Just-in-Time compiled,
interpreted) and distinct trade-offs in terms of resource footprint and
initialization latency~\cite{zhang:ACM:2025:WasmRuntimesSurvey}.
While some existing solutions have attempted to optimize the FaaS paradigm
using WASM~\cite{Gadepalli:Middleware:2020:Sledge,Shillaker:ATC:2020:Faasm},
they typically present two main limitations: (i)~they are highly customized
and tightly coupled to a single execution engine, and (ii)~most of them are
strictly designed for purely stateless workloads.
These limitations directly clash with the reality of edge deployments. Edge
environments are inherently resource-constrained and highly
heterogeneous~\cite{varghese:ACM:2022:EdgeBenchmarking,kimovski:internetcomp:2021:cloud-fog-edge,rausch:hotedge:2020:edge-infrastructure}.
Applications running in this continuum are often stateful and exhibit highly
diverse non-functional requirements, particularly in terms of acceptable
response times, memory footprint, and state access patterns.
Consequently, overcoming the limitations of current WASM FaaS runtimes
requires a paradigm shift to address heterogeneity in terms of hardware and
deployment constraints, and characteristics of the workloads.

To address these challenges, we present \sysname (WebAssembly Stateful
Serverless Platform), a framework explicitly designed to bring stateful
serverless execution to the edge-cloud continuum.
\sysname offers applications a fixed API: developers invoke functions and,
from within those functions, access persistent state through a uniform set of
primitives (\texttt{get} and \texttt{put}).
Underneath this stable interface, \sysname is built around the core principle
of configurability, allowing the system to be customized without requiring any
modifications to the application code.
To accommodate diverse hardware, deployment contexts, and workload
characteristics, \sysname is completely agnostic to both the execution engine
and the storage backend, treating them as fully pluggable components.
System administrators can thus swap the WASM runtime and the datastore to best
fit their specific deployment.
This configurability is motivated by concrete edge scenarios in which the same
application may target nodes with radically different CPU, memory, and
persistence constraints, making a low-footprint interpreter with an in-memory
store preferable on some devices and an AOT runtime with a durable backend
preferable on others.
The framework further supports customizable function lifecycle management and
multi-level caching strategies, enabling execution to be optimized for the
specific non-functional requirements and state access patterns of each
application.

To validate our approach, we implemented a fully functional prototype of
\sysname in Go, currently supporting three distinct WASM runtimes (Wasmtime,
Wazero, and WasmEdge) and two external state backends, Redis and PostgreSQL.
Our experimental evaluation demonstrates that the framework introduces
negligible overhead, exposes radically different performance and resource
profiles across the supported runtime and storage combinations, and remains
competitive against a state-of-the-art WASM serverless framework on
lightweight workloads. Finally, we show that \sysname deploys unchanged on
resource-constrained edge hardware, confirming its suitability across the
edge-cloud continuum.

The remainder of this paper is organized as follows.
\s{background} provides the necessary background on stateful serverless
computing and WASM at the edge, discussing the limitations of related work.
\s{design} introduces the architectural design and the implementation of the
\sysname framework, detailing its modular components, configuration
strategies, and state management approach.
\s{eval} evaluates the system, highlighting the distinct performance and
memory profiles achievable through configuration tuning.
Finally, \s{conclusions} concludes the paper and outlines future research
directions.

%% file: background.tex
\section{Background and Related Work}
\label{sec:background}

In this section, we analyze the challenges of serverless state management at
the edge and the emergence of WASM as a lightweight execution environment,
highlighting gaps in the state-of-the-art.

\fakeparagraph{Stateful Serverless Computing}
The traditional separation of compute and storage in FaaS enables easy
horizontal scaling, but leaves applications without a native mechanism to
persist state across invocations. This has been identified as one of the key
challenges for the future of serverless
computing~\cite{schleier-smith:ACM:2021:ServerlessFuture,wen:ACM:2023:RisePlanetServerless}.
A significant body of work has investigated stateful serverless from adjacent
angles, such as workflow orchestration and strong consistency
guarantees~\cite{jia:SIGOPS:2021:Boki,zhang:OSDI:2020:Beldi}; these concerns
are largely orthogonal to ours. Other systems tackle state more directly:
Cloudburst~\cite{sreekanti:cloudburst}, for instance, addresses the
data-locality problem by co-locating mutable caches with function executors.
The problem becomes especially acute in the edge-cloud continuum. At the
network edge, where bandwidth is limited and latency is critical, the
traditional data-shipping architecture that moves data back and forth to a
central cloud database often incurs network delays exceeding the function's
own execution time. Managed solutions such as Azure Durable Functions or
Cloudflare Durable Objects mitigate this, but they are proprietary,
vendor-locked, and cannot be deployed on custom edge devices.
A few systems target stateful execution at the edge directly.
Enoki~\cite{pfandzelter:Middleware:2023:Enoki} brings stateful functions to
the edge by combining Docker with the
FReD~\cite{pfandzelter:SoftwarePracticeExperience:2023:FRED} key-value store,
while LoLa~\cite{wen:ISPA:2024:StateManagementEdgeServerless} introduces a
state-management abstraction for real-time edge environments, supporting
persistence and migration. However, these platforms rely on heavy container
runtimes (e.g., Docker) ill-suited for strict edge constraints, or lack
abstractions for interchangeable datastores. Consequently, there is a need for
lightweight, environment-agnostic solutions that manage state without
overwhelming edge resources.

\fakeparagraph{WASM in the Edge-Cloud Continuum}
Recent benchmarks show WASM is highly suited for the edge, drastically
reducing image size and memory usage compared to containers, while maintaining
near-native performance~\cite{Liu:ACM:2025:WasmVsDocker}.
The WASM ecosystem features diverse runtimes (e.g., Wasmtime, WasmEdge,
Wazero) with various execution strategies. Ahead-of-Time (AOT) compilers
translate bytecode into native machine code pre-execution, minimizing startup
latency but producing platform-dependent binaries. Just-In-Time (JIT)
compilers dynamically translate code, balancing peak performance and
portability. Interpreters avoid compilation overhead, excelling in
memory-constrained devices at the cost of raw speed.
Several recent works leverage WASM to optimize serverless computing.
Sledge~\cite{Gadepalli:Middleware:2020:Sledge} achieves extreme efficiency at
the edge via AOT compilation and bypassing the OS kernel. However, Sledge is
strictly stateless, lacks state management abstractions, and operates as a
monolithic environment.
Conversely, Faasm~\cite{Shillaker:ATC:2020:Faasm} introduces state via
Faaslets, lightweight WASM units sharing memory regions for zero-copy access.
While excellent for data-intensive tasks, Faasm enforces a customized memory
model tightly coupled to its execution logic, preventing runtime
interchangeability.
Similarly, WasmPulse~\cite{pang:Transactions:2025:WasmPulse} tackles stateful
inference by executing functions as threads in WASM modules, leveraging
lock-free, zero-copy intra-process memory sharing. Like Faasm, it operates as
a monolithic framework tightly integrated with Kubernetes, lacking modular
configurability.

\fakeparagraph{Research Gap}
Two limitations recur across the state-of-the-art. In \emph{state management},
existing WASM serverless systems either ignore state or expose it through
backend-specific mechanisms, offering no abstraction that survives a change of
datastore. In terms of \emph{heterogeneity}, these monolithic systems lock
developers into a single execution engine, lacking the architectural
flexibility required by diverse edge devices.
\sysname fills this gap with a modular architecture for stateful serverless
applications that decouples state management from execution: applications
access state through a fixed, backend-agnostic API, while administrators
tailor the runtime, the datastore, and the execution policies to the specific
constraints of the deployment.

%% file: design.tex
\section{System Design and Implementation}
\label{sec:design}

\begin{figure}[tpb]
  \centering
  \includegraphics[width=0.8\textwidth]{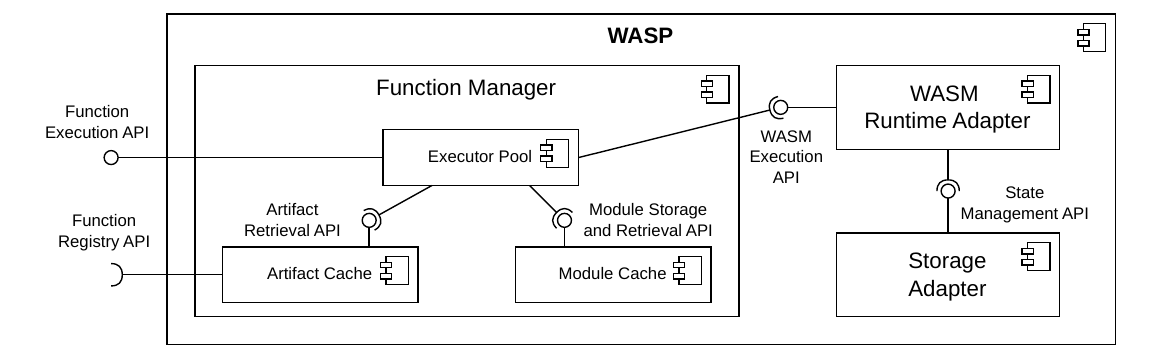}
  \caption{The \sysname framework architecture}
  \label{fig:arch}
\end{figure}

\sysname is designed around strict principles of modularity and separation of
concerns.
\fig{arch} illustrates the component diagram of the framework, highlighting
its external interfaces and internal subsystems.

From an external perspective, \sysname acts as a stateful serverless gateway.
It exposes a simple \texttt{Function Execution API}: given the name of a
target function and a set of input parameters serialized as a byte array
(e.g., a JSON payload), it triggers the execution and returns the computed
result.
During their execution functions are not confined to their inputs: they can
read and persist state through a uniform \texttt{State Management API}.

To fulfill these requests, the framework assumes the existence of an external
\emph{Function Registry}, accessed via the \texttt{Function Registry API},
which acts as a repository of WASM binaries: given a function name, this
registry returns the corresponding \texttt{.wasm} binary.
Functions are assumed to be versioned, allowing developers to seamlessly
update and replace logic at runtime.

From a developer's perspective, the workflow is straightforward: application
logic is compiled into a standard \texttt{.wasm} module using any supported
language toolchain, and then uploaded to the Registry. This entirely decouples
the function development lifecycle from the actual deployment and execution
configuration of \sysname.

Internally, the architecture avoids monolithic execution paradigms and is
instead structured into three primary, decoupled components:
\begin{inparaenum}[(1)]
  \item \emph{Function Manager}, the operational core that serves the
  \texttt{Function Execution API} by scheduling and orchestrating execution
  through an \emph{Executor Pool} of concurrent workers and a multi-level
  cache (\emph{Artifact Cache} and \emph{Module Cache});
  \item \emph{WASM Runtime Adapter}, a generic wrapper around diverse WASM
  runtimes that exposes a uniform \texttt{WASM Execution API}; and
  \item \emph{Storage Adapter}, a wrapper around external datastores (e.g.,
  Redis) that exposes a unified \texttt{State Management API}.
\end{inparaenum}
This decoupling is what makes \sysname configurable: system administrators can
tailor the platform to the specific requirements of the application and the
capabilities of the host device.

\subsection{WASM Runtime Adapter}

A core design philosophy of \sysname is that no single WASM runtime is
universally optimal. Depending on deployment constraints and workload
characteristics, applications may require different performance profiles.
To accommodate this diversity, \sysname integrates different WASM runtimes
through the WASM Runtime Adapter. This integration is a complex software
engineering challenge because runtimes possess radically different internal
memory models and concurrency paradigms (e.g., C/Rust-based engines versus
pure Go-based implementations).
The adapter pattern in \sysname encapsulates these underlying constructs and
exposes a uniform \texttt{WASM Execution API}. This standardizes critical
operations: binary compilation, context instantiation, execution invocation,
and memory boundary traversal.

Beyond accommodating diverse runtimes, applications also benefit from
controlling how each module is prepared for execution. Some workloads, such as
long-running or frequently-invoked functions, amortize an upfront compilation
cost over many invocations, while others prefer the responsiveness of the
runtime's default execution strategy.
\sysname therefore exposes two execution modes.
\begin{inparaenum}[(1)]
  \item A \modeInt mode, in which the binary is loaded as-is and executed
  using the runtime's default strategy, whichever it may be (e.g., JIT on
  Wasmtime, interpretation on WasmEdge).
  \item An \modeAot mode, in which the \texttt{.wasm} binary is explicitly
  compiled to native code at load time, and the resulting module is cached for
  subsequent executions.
\end{inparaenum}
This abstraction provides deployment flexibility: administrators can swap
runtimes and execution modes via a simple configuration flag, tailoring the
CPU-to-memory trade-off without altering a single line of the framework's core
networking or scheduling logic.

To validate this approach, our prototype implementation, built in Go, includes
adapters for three distinct WASM engines chosen to represent different
execution paradigms:
\begin{inparaenum}[(i)]
  \item Wasmtime, a mature C/Rust-based runtime integrated via CGO that
  employs a Just-In-Time (JIT) compiler (Cranelift) as its default execution
  strategy and supports thread-safe shared objects;
  \item Wazero, a zero-dependency, pure-Go runtime that provides AOT
  compilation and enables highly portable, statically compiled binaries; and
  \item WasmEdge, a high-performance C++ runtime that defaults to
  interpretation but supports explicit AOT compilation.
\end{inparaenum}

\subsection{Function Manager and Multi-level Caching}

The Function Manager orchestrates executions through three sub-components,
described below.

\fakeparagraph{Executor Pool}
Drawing inspiration from traditional thread pools, the Executor Pool acts as a
concurrency limiter, managing a finite set of WASM workers to prevent resource
exhaustion on the host node. In our prototype, this is heavily supported by
Go's native concurrency model based on lightweight goroutines and channels,
enabling high-throughput request multiplexing with minimal OS-level thread
overhead.
The pool's behavior is parametrized by the runtime adapter in use, and we
illustrate this with the three runtimes integrated in our prototype.
For runtimes designed for high-concurrency (e.g., Wasmtime, Wazero), it
leverages shared, thread-safe objects. A heavyweight engine is shared across
all executions, and workers create lightweight, ephemeral execution contexts
for each call.
Conversely, for runtimes not designed for concurrent sharing (e.g., WasmEdge),
it maintains a set of complete, pre-allocated execution environments. A
request borrows an entire environment for the duration of the execution and
returns it upon completion. To guarantee security and isolation, the pool
resets instances to a clean state after each use, by clearing the linear
memory and global state to ensure no data leaks across invocations.

\fakeparagraph{Multi-level Caching Strategy}
To mitigate the cold start penalty associated with code retrieval and
compilation, the pool relies on a two-tier caching mechanism.
An Artifact Cache (Layer 1) stores the raw \texttt{.wasm} binaries locally
after downloading them via the \texttt{Function Registry API}. To prevent the
\textit{cache stampede} problem during concurrent invocations of an uncached
function, this cache implements a \emph{singleflight} pattern. This ensures
only one network request is dispatched to the registry, while concurrent
invocations are suspended and subsequently served from the newly populated
cache.

A Module Cache (Layer 2) stores compiled modules in main memory. Once a
function's binary has been loaded from the Artifact Cache (independently of
the \sysname execution mode), it is transformed into a runtime-specific module
ready for execution. This module is then stored in the Module Cache, which
bypasses the often unbounded native caches of different runtimes and is
implemented through a custom, unified Least Recently Used (LRU) Module Cache
across all adapters. For ``warm starts'', the Executor Pool retrieves the
pre-compiled module directly from memory, completely bypassing the module
preparation phase.

\subsection{Unified State API via Storage Adapter}

Maintaining the FaaS paradigm's simplicity requires abstracting how state is
persisted. Embedding database drivers directly within the WASM binaries would
inflate their footprint, compromise security, and tightly couple functions to
specific database technologies.
To resolve this, \sysname implements a Storage Adapter that exposes a unified
\texttt{State Management API} (\texttt{get}, \texttt{put}).

In our prototype, we integrated Redis and PostgreSQL as data stores, utilizing
the \texttt{go-redis} and \texttt{go-pq} clients respectively. The adapter
maps the generic API directly to the respective database commands.

This API is made accessible to the guest WASM sandboxes through secure
\emph{Host Functions}.
Because the WASM specification limits cross-boundary calls to simple numeric
types, \sysname employs a structured memory-sharing protocol. To exchange
complex data (e.g., JSON payloads), the guest allocates a block in its linear
memory, writes the payload, and passes the memory \texttt{offset} and
\texttt{length} to the Host Function.
The framework safely reads this exact byte sequence, processes the state
operation through the Storage Adapter, and writes the response back into the
guest's memory.
Notably, these low-level memory management operations are completely
abstracted from the application developer. \sysname provides a dedicated SDK
that exposes idiomatic state access methods in the target programming
language; in our prototype, serverless functions are written in Go and
compiled to \texttt{.wasm} via TinyGo. From the developer's perspective,
reading and writing state amounts to a single function call in the source
code; all interactions with the Host Functions and linear memory are handled
internally by the SDK.

%% file: eval.tex
\section{Evaluation}
\label{sec:eval}

The aim of our evaluation is to verify whether \sysname lives up to the
promise of configurability stated in \s{intro}: the runtime, the storage backend,
and the deployment target can be varied to fit heterogeneous workloads and
hardware, without modifying application code. We first ask whether the
framework effectively delivers this configurability across its pluggable
components, both in terms of the performance profiles it exposes and the
engineering cost of integration (RQ1); we then measure the runtime overhead
introduced by the framework's abstraction layer (RQ2), and compare \sysname's
performance against a state-of-the-art alternative (RQ3); finally, we verify
the framework's viability on resource-constrained edge hardware (RQ4).

\fakeparagraph{RQ1}
Is \sysname able to exploit the heterogeneity of different WASM runtimes and
data stores to support adaptable performance and resource profiles, and at
what integration cost?

\fakeparagraph{RQ2}
What is the overhead introduced by \sysname relative to standalone runtimes
and native execution?

\fakeparagraph{RQ3}
How does \sysname's performance compare to that of a state-of-the-art
reference framework?

\fakeparagraph{RQ4}
Can \sysname be deployed on resource-constrained edge devices, beyond the
cloud-class hardware?

\subsection{Experiment Setup}
\label{sec:eval:setup}

\sysname is implemented in Go (v1.25). All experiments were conducted on a
server-class machine equipped with a 16-core AMD Ryzen 9 9950X processor (32
threads) and 64\,GB of DDR5 RAM, running Fedora Server 42. Experiments on an
edge-class device were conducted on a Raspberry Pi 3B+ (BCM2837B0, 4 ARMv8
cores, 1\,GB RAM), running Debian GNU/Linux 13 (Trixie).
We integrate \sysname with three WASM runtimes through their respective Go
APIs: Wasmtime (v44.0.1, default JIT), WasmEdge (v0.16.3, default
interpretation), and Wazero (v1.11.0, default AOT). We compare each
\sysname-runtime combination against the corresponding runtime invoked
standalone via its native CLI. As an additional reference, we also report
native Go execution. The Storage Adapter is instantiated with Redis (v8.6.2)
and PostgreSQL (v18.4) as backends. WASM modules are compiled from Go sources
using TinyGo (v0.41.1, \texttt{-target=wasi -buildmode=c-shared}). We evaluate
both the \modeAot and \modeInt modes, described in \s{design}.

To answer RQ1 and RQ2, we measure execution time and peak RSS memory across
three workloads. The two computational ones are adopted from the application
suite shipped with Sledge~\cite{Gadepalli:Middleware:2020:Sledge}, the
state-of-the-art framework we compare against in
RQ3:\footnote{\url{https://github.com/gwsystems/wasm_apps}}
\texttt{Fibonacci(10)}, which recursively computes the 10th Fibonacci number
as a lightweight, call-intensive CPU-bound function (averaged over 100 runs),
and \texttt{Hash(30000)}, which applies the EJB hash algorithm over 30{,}000
iterations as a sustained, loop-intensive CPU-bound workload (averaged over 10
runs). The third workload is a \texttt{Stateful} function that performs a
\texttt{set} followed by a \texttt{get} on the same key-value pair via the
Storage Adapter (averaged over 100 runs). Each test distinguishes between
\textit{Cold} execution (the first run, no cached modules) and \textit{Warm}
execution (a subsequent run served from the Module Cache).
To answer RQ3, we compare \sysname against Sledge, a state-of-the-art WASM
serverless framework highly optimized for edge environments. We run an HTTP
benchmark measuring throughput and latency over 10 repetitions of 10-second
runs at varying request rates. \sysname is configured with 32 workers (one per
core); Sledge uses 30 workers, 1 HTTP listener, and 1 worker controller. The
HTTP frontend of \sysname is implemented using \texttt{fasthttp}. For
fairness, all \sysname runtimes are set to \modeAot mode, since Sledge
requires pre-compiled modules via \texttt{aWsm}.
To answer RQ4, we deploy \sysname on the edge-class device and replicate the
RQ1 workloads.

\subsection{Results}

\tab{benchmarks} reports the cold-start and warm-start execution times and the
peak RSS for \sysname across the six runtime/mode combinations, together with
the corresponding standalone runtimes and native Go execution as a baseline.
For stateful workloads, the table additionally reports the same metrics for
the \texttt{Stateful} function under the two storage backends.

\begin{table}[t]
  \centering
  \scriptsize
  \renewcommand{\arraystretch}{1.0}
  \setlength{\tabcolsep}{4pt}
  \begin{tabular}{@{}l l rrr rrr@{}}
    \toprule
    \textbf{Runtime}                                                         &
    \textbf{Config.}                                                         &
    \textbf{Cold}                                                            &
    \textbf{Warm}                                                            &
    \textbf{RSS}                                                             &
    \textbf{Cold}                                                            &
    \textbf{Warm}                                                            &
    \textbf{RSS}
    \\
    \cmidrule(lr){3-5} \cmidrule(lr){6-8}                                    &
                                                                             &
    \multicolumn{3}{c}{\textbf{Fibonacci}}
                                                                             &
    \multicolumn{3}{c}{\textbf{Hash}}
    \\
    \midrule
    \multirow{4}{*}{\textit{Wasmtime}}                                       &
    \sysname-AOT                                                             &
    6.15                                                                     &
    0.08                                                                     &
    48.94                                                                    &
    180.59                                                                   &
    172.57                                                                   &
    48.59                                                                      \\
                                                                             &
    \sysname-JIT (D)                                                         &
    6.06                                                                     &
    0.08                                                                     &
    48.88                                                                    &
    177.88                                                                   &
    172.56                                                                   &
    48.80                                                                      \\
                                                                             &
    Standalone-AOT                                                           &
    13.97                                                                    &
    --                                                                       &
    19.84                                                                    &
    187.73                                                                   &
    --                                                                       &
    19.24                                                                      \\
                                                                             &
    Standalone-JIT (D)                                                       &
    3.06                                                                     &
    --                                                                       &
    19.21                                                                    &
    177.49                                                                   &
    --                                                                       &
    19.92                                                                      \\
    \midrule
    \multirow{4}{*}{\textit{WasmEdge}}                                       &
    \sysname-AOT                                                             &
    317.25                                                                   &
    0.06                                                                     &
    68.66                                                                    &
    448.21                                                                   &
    129.55                                                                   &
    68.75                                                                      \\
                                                                             &
    \sysname-Interp.\ (D)                                                    &
    4.21                                                                     &
    3.47                                                                     &
    29.01                                                                    &
    11{,}783.39                                                              &
    11{,}759.89                                                              &
    29.26                                                                      \\
                                                                             &
    Standalone-AOT                                                           &
    573.95                                                                   &
    --                                                                       &
    69.48                                                                    &
    701.70                                                                   &
    --                                                                       &
    69.47
    \\
                                                                             &
    Standalone-Interp.\ (D)                                                  &
    8.28                                                                     &
    --                                                                       &
    24.39                                                                    &
    11{,}766.90                                                              &
    --                                                                       &
    24.48                                                                      \\
    \midrule
    \multirow{3}{*}{\textit{Wazero}}                                         &
    \sysname-AOT                                                             &
    9.86                                                                     &
    0.15                                                                     &
    33.27                                                                    &
    181.65                                                                   &
    172.50                                                                   &
    33.37                                                                      \\
                                                                             &
    \sysname-AOT (D)                                                         &
    9.25                                                                     &
    0.14                                                                     &
    33.27                                                                    &
    181.47                                                                   &
    172.65                                                                   &
    33.06                                                                      \\
                                                                             &
    Standalone-AOT (D)                                                       &
    23.80                                                                    &
    --                                                                       &
    16.52                                                                    &
    198.93                                                                   &
    --                                                                       &
    16.48                                                                      \\
    \midrule
    \multicolumn{2}{l}{\textit{Native (Go)}}                                 &
    0.59                                                                     &
    --                                                                       &
    1.79                                                                     &
    133.03                                                                   &
    --                                                                       &
    1.87
    \\
    \midrule
    \midrule
                                                                             &
                                                                             &
    \multicolumn{3}{c}{\textbf{Redis}}
                                                                             &
    \multicolumn{3}{c}{\textbf{PostgreSQL}}
    \\
    \cmidrule(lr){3-5} \cmidrule(lr){6-8} \multirow{2}{*}{\textit{Wasmtime}} &
    \sysname-AOT                                                             &
    38.35                                                                    &
    0.19                                                                     &
    141.52                                                                   &
    40.22                                                                    &
    1.49                                                                     &
    140.84                                                                     \\
                                                                             &
    \sysname-JIT (D)                                                         &
    37.41                                                                    &
    0.34                                                                     &
    138.96                                                                   &
    38.86                                                                    &
    1.46                                                                     &
    138.21                                                                     \\
    \midrule
    \multirow{2}{*}{\textit{WasmEdge}}                                       &
    \sysname-AOT                                                             &
    7{,}906.37                                                               &
    0.21                                                                     &
    180.45                                                                   &
    7{,}919.31                                                               &
    16.60                                                                    &
    179.57                                                                     \\
                                                                             &
    \sysname-Interp.\ (D)                                                    &
    13.29                                                                    &
    6.23                                                                     &
    41.10                                                                    &
    14.82                                                                    &
    7.25                                                                     &
    40.42                                                                      \\
    \midrule
    \multirow{2}{*}{\textit{Wazero}}                                         &
    \sysname-AOT                                                             &
    130.80                                                                   &
    0.35                                                                     &
    59.67                                                                    &
    134.14                                                                   &
    2.17                                                                     &
    59.15                                                                      \\
                                                                             &
    \sysname-AOT (D)                                                         &
    131.15                                                                   &
    0.35                                                                     &
    59.70                                                                    &
    134.36                                                                   &
    2.17                                                                     &
    59.06
    \\
    \bottomrule
  \end{tabular}
  \caption{Execution times (Cold, Warm) in milliseconds [ms] and peak RSS in
    Megabytes [MB] across benchmarks. ``Standalone'' refers to the runtime
    invoked via its native CLI without \sysname; ``(D)'' marks each runtime's
    default execution mode.}
  \label{tab:benchmarks}
\end{table}

\fakeparagraph{Exploiting Heterogeneity (RQ1)}
\sysname highlights a wide spectrum of performance and resource profiles
across the underlying runtimes and execution modes. The AOT compilation mode
(\modeAot) achieves near-native execution times on \textit{Warm} runs, but
caching heavily dictates AOT performance: during \textit{Cold} starts,
WasmEdge \modeAot incurs prolonged compilation times (e.g., 317\,ms on
Fibonacci) due to its LLVM-based backend. Conversely, in \modeInt the engine
relies on its native execution strategy, which initiates lightweight execution
faster than \modeAot but can degrade sharply on computationally heavy tasks:
WasmEdge's interpreter requires nearly 12\,s for Hash. Wazero is an exception,
as its default running mode, \textit{AOT}, renders the two \sysname execution
modes effectively equivalent in our measurements.

Memory usage exhibits a similar spread, from 29\,MB (WasmEdge
\textit{Interp.}) to 69\,MB (WasmEdge \modeAot), reflecting the cost of
keeping compiled modules in memory. The Stateful function amplifies this
effect: the AOT module inflates peak RSS by $3$--$4\times$ over the
corresponding interpreted configuration (e.g., 41\,MB vs.\ 180\,MB on
WasmEdge), as our \sysname SDK needs to be fully embedded in the
\texttt{.wasm} binary to support the integration of Host Functions for storage
access.

The same configurability extends to the two integrated storage backends, Redis
and PostgreSQL. The choice of backend is largely transparent during Cold
Start, where latency is dominated by the WASM runtime initialization overhead.
The architectural differences emerge in Warm execution: a warm Stateful
execution on Wasmtime-\modeAot takes 0.19\,ms with in-memory Redis, compared
to 1.49\,ms with PostgreSQL. PostgreSQL is inherently slower due to disk
persistence and relational overhead, yet remains within the millisecond range.

\fakeparagraph{Integration Cost (RQ1)}
The performance variety described above is made available through \sysname's
adapter pattern, at a low engineering cost. Integrating a new WASM runtime
required between 245 and 317 lines of Go code (\textit{Wazero}: 245,
\textit{Wasmtime}: 263, \textit{WasmEdge}: 317), most of which wrap
engine-specific concurrency primitives and linear memory boundary handling.
Similarly, the two storage adapters required 53 (Redis) and 76 (PostgreSQL)
lines respectively. In both cases, the integration work is entirely confined
to the adapter layer.

\fakeparagraph{Framework Overhead (RQ2)}
As established in the
literature~\cite{Talu:advancedengscience:2025:WasmRuntimes,
  khelifa:GIIS:2024:WasmAtEdgeAndAI}, WASM execution introduces performance
slowdown and memory inflation relative to native code; our experiments confirm
this, with native Go using 1.8\,MB of peak RSS against tens of MB for all
runtime-based configurations.

Focusing on the cost of the framework itself, \sysname's cold-start latency is
comparable to, and in several cases lower than, the corresponding standalone
runtime. For example, on Fibonacci, \sysname-Wasmtime-AOT runs in 6.2\,ms
compared to 14.0\,ms for the standalone runtime, since the framework amortizes
runtime initialization across invocations whereas the standalone CLI
re-instantiates the engine on each call. On Hash, the two converge within 4\%
(180.6 vs.\ 187.7\,ms), showing that the framework adds no measurable cost on
the compute-bound portion of the execution. The Module Cache yields up to
two-order-of-magnitude speedup on warm starts (e.g., from 6.2\,ms to 0.08\,ms
on Wasmtime-AOT), bringing warm execution within a small constant factor of
native Go.
The price paid for this flexibility is in memory. \sysname's peak RSS exceeds
the corresponding standalone runtime by 10--50\,MB, due to the worker pool
maintaining live module instances and the HTTP front-end. On stateful
workloads, the gap widens further when modules are compiled AOT, for the
reasons discussed above. In absolute terms, however, \sysname remains within
tens of megabytes.

\begin{figure*}[tbp]
  \centering
  \includegraphics[width=0.90\textwidth]{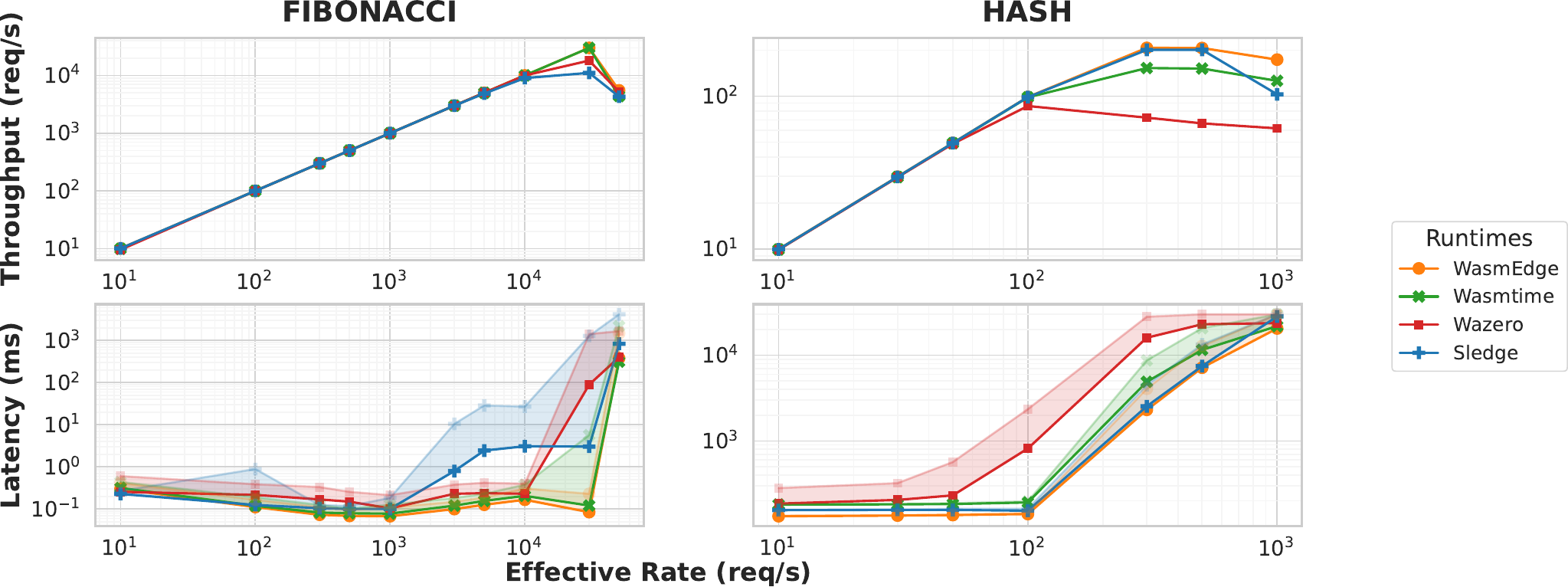}
  \caption{Throughput versus request rate and latency distribution for
    Fibonacci and Hash. Latency reports the median (p50, line) and the area up
    to the 90th percentile.}
  \label{fig:throughput_latency}
\end{figure*}

\fakeparagraph{Comparison with Sledge (RQ3)}
\fig{throughput_latency} shows that \sysname and Sledge exhibit different
behaviors depending on the workload.
For lightweight operations (Fibonacci), our framework scales uniformly better
than Sledge. It achieves higher peak throughput and maintains a lower, more
stable 90th-percentile latency even under severe load. The only exception is
WasmEdge, which hits a saturation ceiling dictated by the limited size of its
reusable environment pool. This advantage stems from \sysname's HTTP layer, which distributes requests across all cores via \texttt{fasthttp} and goroutine multiplexing, whereas Sledge funnels them through a single listener core.

Conversely, heavyweight, CPU-bound workloads (Hash) expose a different trend.
Under these conditions, Sledge's architectural choices yield better overall
stability. While our framework with WasmEdge manages to maintain throughput
and latency comparable to Sledge, the other runtimes experience noticeable
performance degradation. This inversion is rooted in the scheduling
architecture. Our prototype relies on a simple FIFO dispatcher that delegates
thread management to the underlying OS scheduler. During prolonged, intensive
computations, the OS aggressively preempts threads, introducing heavy
context-switching penalties. Pure Go implementations like Wazero suffer the
worst regressions from this context switching across the pool structure.
Sledge mitigates this by employing a highly customized, unguided scheduler
designed specifically to minimize context switches during long-running tasks.

\fakeparagraph{Edge Portability (RQ4)}
The framework runs without modification on all three runtimes, confirming the
portability promised by its architectural design. As expected, absolute
performance degrades compared to the server-class machine: on Fibonacci,
\sysname-Wasmtime-AOT moves from 6.2\,ms cold and 0.08\,ms warm on the server
to 138\,ms cold and 1.4\,ms warm on the Pi. The Stateful function shows the
same scaling behavior, with \sysname-Wasmtime-AOT on Redis going from 38\,ms /
0.19\,ms (cold/warm) to 1.82\,s / 3.8\,ms. Despite the two-order-of-magnitude
slowdown on cold starts, peak RSS remains well within the Pi's 1\,GB envelope
(between 28 and 277\,MB across all configurations, with the upper bound
reached by \sysname-WasmEdge-AOT on the stateful workload).

\subsection{Discussion}

Across the four research questions, \sysname does not introduce a
one-size-fits-all configuration: different runtimes, modes, and storage
backends yield radically different performance and memory profiles (RQ1), and
the framework's abstraction layer preserves these differences while adding
negligible overhead (RQ2). When compared to a highly customized
state-of-the-art framework (RQ3), \sysname remains competitive: outperforming
Sledge on lightweight workloads while staying within range on heavyweight
ones, despite a deliberately naive scheduler. Crucially, the same framework
deploys unchanged on a Raspberry Pi (RQ4), confirming that this configurability
is not confined to cloud-class hardware. Taken together, these results
validate the central premise of \sysname: that stateful WASM serverless can be
both portable across heterogeneous deployment targets and configurable to
workload-specific requirements, without requiring application code changes.

%% file: conclusions.tex
\section{Conclusions}
\label{sec:conclusions}

In this paper, we presented \sysname, a configurable framework that brings
stateful serverless computing to the heterogeneous edge-cloud continuum.
Structured around strictly decoupled components (a Function Manager, a WASM
Runtime Adapter, and a Storage Adapter), \sysname overcomes the rigidity of
existing WASM-based serverless solutions. Our prototype confirms that this
modularity comes at negligible runtime cost, while letting administrators tune
the execution engine, the datastore, and the caching policies, and deploy the
same framework from server-class machines down to resource-constrained edge
devices. As a result, \sysname exposes distinct latency and memory profiles,
allowing the execution environment to be precisely tailored to the
requirements of each application and the constraints of its deployment target.

Future work will extend \sysname with additional runtimes and datastores, a
richer state API, and smarter scheduling, and build a full serverless platform
on top of it.